\documentclass[reprint,prd,nofootinbib]{revtex4-1}

\usepackage{color}
\usepackage[dvipsnames]{xcolor}
\usepackage{ulem}

\usepackage{graphicx}
\usepackage{amssymb}
\usepackage{amsmath}
\usepackage{epstopdf}
\usepackage{subcaption}
\usepackage{bm}
\usepackage{hyperref}

\def\bra#1{\langle#1\vert}
\def\ket#1{\vert#1\rangle}

\def\q{{\bm q}}
\def\l{{\boldsymbol l}}
\def\k{{\boldsymbol k}}

\def\p{{\boldsymbol p}}
\def\x{{\boldsymbol x}}
\def\y{{\boldsymbol y}}

\def\r{{\boldsymbol r}}

\newcommand{\rme}{{\rm e}}

\newcommand{\nn}{\nonumber\\ }
\newcommand{\beq}{\begin{eqnarray}}
\newcommand{\eeq}{\end{eqnarray}}

\begin{document}

\title{Radiative corrections to the jet quenching parameter in dilute and dense media}
\author{Jean-Paul Blaizot}
\email{jean-paul.blaizot@ipht.fr}
\affiliation{Institut de Physique Th\'eorique, Universit\'e Paris Saclay, CEA, CNRS, F-91191
Gif-sur-Yvette, France}
\author{Fabio Dominguez}
\email{fabio.dominguez@usc.es}
\affiliation{Instituto Galego de F\'isica de Altas Enerx\'ias IGFAE, Universidade de Santiago de Compostela, Santiago de Compostela, 15782 Galicia, Spain}

\begin{abstract}
We extend the calculation of radiative corrections to momentum broadening in a QCD medium to include processes where single scattering dominates over multiple soft scatterings. Our analysis focusses on processes that are enhanced by a potentially large double logarithmic contribution. We demonstrate that such double logarithmic contributions emerge in calculations of momentum broadening in dilute systems where only single scattering occurs. We show how the different regimes, multiple versus single scattering, constrain different parts of the phase space.
\end{abstract}

\maketitle

\section{Introduction}

While the suppression of high energy particles in heavy ion collisions has been regarded as one of the main measurements to claim the formation of a quark-gluon plasma, it is generally recognized that in order to be able to extract precise information about the properties of the plasma from such measurements, a better understanding of the interactions between jets and plasma is necessary. The wealth of data from the LHC heavy ion program has shown that there are many effects that should be taken into account in any theoretical model aiming at  accurate predictions. In particular, the fact that jets evolve and become multi-particle systems while traversing the plasma poses a major difficulty to have a proper description of the interaction between probe and medium.

Some of the first attempts to go beyond the picture of independent emissions and independent propagation of the partons in the shower, while including effects of coherence, are the calculations of radiative corrections to some inclusive observables like transverse momentum broadening or energy loss \cite{Liou:2013qya,Blaizot:2013vha,Blaizot:2014bha}. The main ideas in those calculations is to consider the emission of a soft gluon, which is not measured but can interact with the medium, and study how this process affects the inclusive observable that one considers. In principle, such radiative corrections are suppressed by powers of the coupling constant, but they have been shown to receive a double logarithmic enhancement which makes them potentially large.

So far, the calculations that have been successful at estimating these double logarithmically enhanced contributions are all set in a multiple scattering framework where the length of the medium is much larger than the mean free path, making edge effects negligible. Under those conditions, it is common practice to identify the relevant momentum scale, which is used as an upper cutoff for transverse momentum integrations, with $\hat q L$ -- the jet quenching parameter ($\hat q$) times the length ($L$) of the medium. This identification has for consequence that the two logarithms from the radiative corrections become controlled by the same scale. In Refs.~\cite{Iancu:2014sha,Mueller:2016xoc} however, it was correctly pointed out that the momentum scale regulating the collinear divergence can be larger than $\hat q L$, in which case one of the logarithms appearing in the radiative corrections is given by an integration over the formation time of the radiated gluons and its phase space is directly limited by the length of the medium.

The problem with the approach just mentioned is that when the transverse momentum scale is allowed to run to higher values, it opens up a phase space region where the emissions are dominated by single scattering mechanisms for which the approximations that are usually made in the context of a multiple scattering formalism are no longer valid. In the multiple scattering formalism it is always assumed that the medium is sufficiently long and the main contribution to medium radiation comes from gluons which have a formation time much shorter than the length of the medium. Once we move into the region where the dynamics is dominated by a single scattering, longer formation times should be considered and  edge effects become important. In fact, as we shall show by explicit calculation,  even though formation times can be typically very large, the double logarithmically enhanced contribution remains dominated by the region of phase space where the formation time is shorter than the length of the medium.

These radiative corrections have been argued to be a universal feature of the interaction between probe and medium, leading to an interpretation as a renormalization of the jet quenching parameter, at least to leading logarithmic accuracy \cite{Blaizot:2014bha}. In this setup, the evolution equation for $\hat q$ in the double logarithmic approximation resembles the double logarithmic limit of DGLAP, which is consistent with the observation that $\hat q$ is directly related to a gluon distribution, as noted in \cite{Baier:1996sk}.  
Such a relation is not so clear, though. It has been established only to leading order, in a simplified model with independent scatterings centers, and with several assumptions which do not allow to clarify in particular the value of the energy fraction $x$ at which the distribution should be evaluated. Thus, we cannot just rely on this relation to find the proper double logarithmic region with its respective evolution variables, as attempted for instance in \cite{CasalderreySolana:2007sw}. An explicit calculation is necessary.

The main purpose of the present paper is to present such an explicit calculation. In doing so, we shall also pursue another goal, that of clarifying why all calculations that focus on single scattering processes apparently fail to exhibit the double logarithmic corrections. Indeed,  we shall show how the double logarithmic radiative correction, that has been found in the multiple scattering setting, also emerges in the case where single scattering mechanisms dominate the dynamics, and we shall identify the proper variables that control such a correction.

The outline of the paper is as follows. The next section reviews briefly some of the standard (leading order) approaches to the calculation of momentum broadening and the jet quenching parameter, illustrating more explicitly some of the connections alluded to in this introduction. The following section, Sect.~\ref{radiativecorrection}, recalls the main features of the leading radiative correction, and in particular the origin of the double logarithmic contribution. The main new results are contained in Sect.~\ref{dilutecase}. There we present the results of the calculations of the radiative correction in the regime dominated by single scattering processes. Finally Sect.~\ref{finitesize} discusses finite size effects on the double logarithmic correction, and Sect.~\ref{conclusions}  summarizes the conclusions.  The Appendix contains technical details, in particular on the calculation of the radiative correction in the single scattering regime.

\section{Transverse momentum broadening and average momentum transferred}

The amount of transverse momentum that a particle picks up while traversing a medium has long been regarded as one of the simplest observables that can provide information about the properties of this medium. This, together with the fact that there is a close relation between momentum broadening and radiative energy loss, has led to numerous studies about the average transverse momentum transfer per unit length, also known as the jet quenching parameter $\hat q$. Before getting into the details of radiative corrections, we shall recall in this section the basis for the leading order treatment of transverse momentum broadening and the definition of the jet quenching parameter. Several approaches exist in the literature, which  essentially capture the same physics, albeit in different languages. 
In this section, we shall attempt to clarify what are the approximations and assumptions that come into play at every stage of such calculations.

\subsection{Diffusion in momentum space}
The most straightforward approach to  transverse momentum broadening  models the propagation of a hard parton as that of a particle undergoing   independent scatterings with medium constituents, with typically a differential Coulomb-like cross-section proportional to\footnote{Other models for the interaction potential which include Debye-screening and other effects more accurately have been tried in the literature. For the logarithmic accuracy we are interested in, this simple model with the respective cutoff at the Debye scale is good enough.}
\begin{equation}\label{Vdeq}
V(\q)=\frac{nN_cg^4}{\q^4},
\end{equation}
such that the average momentum transfer per unit length takes the form
\begin{equation}
\hat q=\int_\q \q^2V(\q).\label{avpperp}
\end{equation}
In these equations, $n$ is the density of scattering centers, $N_c$ the number of colors, $g$ the gauge coupling, and $\q$ the momentum transfer in a collision. The integral in Eq.~(\ref{Vdeq}) is clearly logarithmically divergent. In the infrared, a natural   cutoff is provided by Debye screening. In the ultraviolet, we note that the  shape of the distribution in Eq.~(\ref{Vdeq}) gives much weight to rare events that drive the average momentum transfer upwards. In such circumstances, a more meaningful quantity to characterize the distribution is not its average, but rather its median, or a \emph{typical} momentum transfer \cite{Baier:1996sk,Arnold:2009mr}. We may obtain such a typical momentum transfer by simply putting explicitly an ultraviolet cutoff $\p^2$ in the integral   (\ref{avpperp}), which makes $\hat q$ a scale dependent object. One then gets
\begin{equation}\label{hatqdef}
\hat q(\p^2)=\frac{g^4N_cn}{4\pi}\ln\frac{\p^2}{m_D^2},
\end{equation}
with $m_D$ the Debye  screening mass of the medium.\\

This picture of independent scatterings is consistent with calculations in which the hard parton propagates eikonally (in the $+$ direction) through a random background gauge field whose only relevant component is $A^-(x^+,\x)$ which is independent of $x^-$.\footnote{We use light-cone coordinates with $x^\pm=(x^0\pm x^3)/\sqrt{2}$.} The fluctuations of the random background field are assumed to be Gaussian, with the two-point function given by
\begin{equation}\label{average}
\left\langle A^-_a(x^+,\x)A^-_b(y^+,\y)\right\rangle=\delta_{ab}n\delta(x^+-y^+)\gamma(\x-\y),
\end{equation}
where $n$ is, as in Eq.~(\ref{Vdeq}),  the density of color charges and $\gamma(\x)$ is proportional to the Fourier transform of $V(\q)$ in Eq.~(\ref{Vdeq}), viz. 
\begin{equation}\label{defgamma}
\gamma(\x)=g^2\int_\q \frac{e^{i\q\cdot\x}}{\q^4}.
\end{equation}
The information about the medium with which the hard parton interacts is entirely contained in this correlator (\ref{average}). Note that this fluctuating background field plays the same role as what are referred to as ``Glauber gluons'' in the context of the soft collinear effective theory \cite{Idilbi:2008vm}.\\

There are two aspects of such calculations of momentum broadening that are worth emphasizing. First, the fact that the fields are independent of $x^-$ implies that there is no transfer of energy (the $+$ component of the hard parton momentum  is conserved). This is just a consequence of taking the energy of the initial parton arbitrarily large, which validates the use of the eikonal approximation. And second, the correlations between field fluctutations are local in light-cone time, as reflected in the delta function $\delta(x^+-y^+)$ in Eq.~(\ref{average}), which supposes that the interactions of the hard parton with the medium constituents occur on a time-scale much smaller than all the other time-scales involved in the process. As we shall see in the next subsection this has an impact on the connection of the jet quenching parameter to a gluon distribution. 

\subsection{$\hat q$ as a gluon distribution}

It was first observed in \cite{Baier:1996sk} that $\hat q$ is proportional to the gluon distribution function of the target, from an operator perspective. This was derived following the formalism developed for twist-4 nuclear interactions in \cite{Luo:1993ui}, where \textit{only one scattering} with the medium is taken into account. The key observation is that the cross section for one scattering is proportional to the Fourier transform of the expectation value of a bilocal product of field operators in a hadronic state.

By taking Eq.~(\ref{avpperp}) as starting point, one can easily obtain  such a relation. We have (schematically) 
\begin{widetext}
\begin{align}
\hat q &\propto g^2\int_\q \q^2\int\; d^2\x\, d^2\y\; e^{-i\q\cdot(\x-\y)}\gamma(\x-\y), \nn
&\propto g^2\int_\q\int\; d^2\x\, d^2\y\, dx^+\, dy^+ e^{-i\q\cdot(\x-\y)-iq^-(x^+-y^+)}\partial_\x^i\partial_\y^i\left\langle A_a^-(x^+,\x)A_a^-(y^+,\y)\right\rangle, \nn
&\propto g^2\int_\q\int\; d^2\x\, d^2\y\, dx^+\, dy^+ e^{-i\q\cdot(\x-\y)-iq^-(x^+-y^+)}\;\left\langle F_a^{i-}(x^+,\x)F_a^{i-}(y^+,\y)\right\rangle, \label{qhatgluondis}
\end{align}
\end{widetext}
where  we have made use of the fact that, in the particular light-cone gauge used here, the only relevant component of the background gauge field is the minus component (as we have mentioned earlier). The last line of Eq.~(\ref{qhatgluondis}) clearly shows the relation with the Fourier transform of the average of two field-strength operators, which yields directly the gluon distribution function $xG(x)$, see e.g. \cite{Collins:1981uw}. The derivation above points to one limitation  of the present description: the assumption that the correlation function (\ref{average}) is local in $x^+$ implies that the $q^-$-dependence drops out in Eq.~(\ref{qhatgluondis}) after taking the average, so that the resulting gluon  distribution carries no information about the value of $x$ (energy, or $q^-$, fraction) that is being probed in the target. Thus we expect this kind of description to apply only in regions of moderately small values of $x$, where the gluon distribution $xG(x)$  is expected to have a very weak dependence on $x$ \cite{Kovchegov:1998bi}.\\

Another perspective on this issue comes from  considering the process in which the initial parton is created. When referring to the transverse momentum broadening of a particle it should be kept in mind that the measured particle comes from the fragmentation of a high-energy parton created in the collision,  which subsequently interacts with the created medium. In principle, the process of creation of the initial parton should be taken into account in the calculation, and such process could interfere with the rescatterings responsible for the broadening. Nevertheless, it is always assumed that the initial parton has such a high energy that the process by which it is created is sensitive only to very small length and time scales and therefore can be factored out from the rest of the calculation. Several attempts at understanding this issue, namely how a newly created particle interacts with the medium constituents, have been done for a DIS-like process,  this being the simplest setup for creating a  high-energy parton.

\begin{figure}
\centering
\includegraphics[width=\columnwidth]{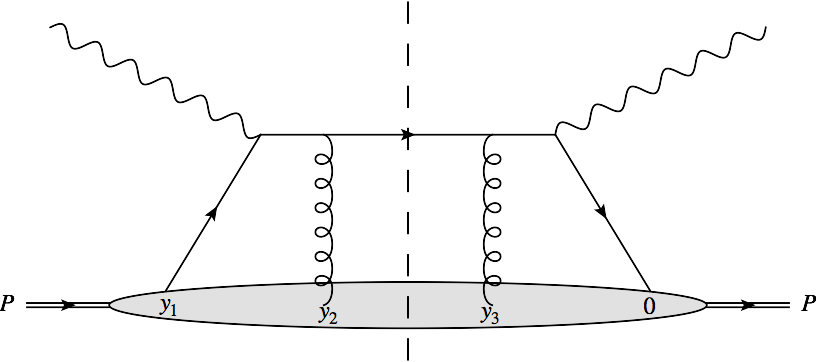}
\caption{DIS-like process in which a quark is created and then rescatters. The amplitude is on the left side and the conjugate amplitude on the right side of the vertical dashed line representing the cut. The independent energy variables attached to the lower part of the diagrams are, from left to right, $x_1 P^-$, $x_2 P^-$, $(x_3-x_2)P^-$.}
\label{figDIS}
\end{figure}

Consider then the case where a high-energy quark is created in a DIS-like process and then rescatters once with the medium in which it propagates, as shown in Fig. \ref{figDIS} (see also \cite{Mueller:2016xoc} for a similar analysis in a related context). The relevant matrix element entering the calculation of the cross section for the process corresponds to the lower part of the diagram in Fig.~\ref{figDIS} and takes the form
\begin{widetext}
\begin{align}
T_A(x_1,x_2,x_3,\q)&=\int\frac{dy_1^+}{2\pi}\frac{dy_2^+}{2\pi}\frac{dy_3^+}{2\pi}\frac{d^2\y}{(2\pi)^2}\;e^{ix_1P^-y_1^+}e^{ix_2P^-(y_2^+-y_3^+)}e^{ix_3P^-y_3^+}e^{i\q\cdot\y}\nn
&\qquad\times\theta(y_2^+-y_1^+)\theta(y_3^+)\langle P|\bar\psi(0)\gamma^+A^-(y_3^+,0)A^-(y_2^+,\y)\psi(y_1^+)|P\rangle,
\end{align}
\end{widetext}
where $|P\rangle$ denotes the target initial state. When calculating the average $\q^2$ the gauge field operators above are replaced by field-strength operators in the same manner as in (\ref{qhatgluondis}).

The assumption that the production of the quark can be factored out from the rest of the process is equivalent to assuming that the matrix element above can be factored into a matrix element for the quark operators times a matrix element for the gauge field operators. This is tantamount to assuming that  the correlations between the emission process and the interaction with the rest of the target are negligible, a reasonable assumption when the energy of the hard parton is sufficiently high\footnote{This is in particular guaranteed in the DIS on a nucleus when the parton is created in the interaction of the virtual photon with one nucleon, and subsequently interact with another nucleon of the nucleus.}. In that case, the relevant matrix element takes the form
\begin{widetext}
\begin{align}
\int_\q \q^2T_A(x_1,x_2,x_3,\q)&=\int\frac{dy_1^+}{2\pi}\frac{dy_2^+}{2\pi}\frac{dy_3^+}{2\pi}\;e^{ix_1P^-y_1^+}e^{ix_2P^-(y_2^+-y_3^+)}e^{ix_3P^-y_3^+}\theta(y_2^+-y_1^+)\theta(y_3^+)\nn
&\qquad\times\langle P|\bar\psi(0)\gamma^+\psi(y_1^+)|P\rangle\bra{P} F^{i-}(y_3^+)F^{i-}(y_2^+)\ket{P}.
\end{align}
\end{widetext}
The equation above would become a product of parton distributions in the absence of the $\theta$-functions enforcing a particular time ordering. But these $\theta$-functions play no role in the relevant regime. Indeed, the production of the hard quark is very localized in space, therefore the integration over $y_1$ has a very small support and effectively decouples from the gluon part of the matrix element, yielding a quark distribution function $f(x_1)$. If $x_2$ is not too small, the remaining integrals are only sensitive to the region where $y_3^+-y_2^+$ is small so that the constraints in $y_2^+, y_3^+$ individually are not important. Under such conditions,  it is easy to see that
\begin{equation}
\int_\q \q^2T_A(x_1,x_2,x_3,\q)\sim f(x_1)\; x_2 G(x_2),
\end{equation}
with $x_2 G(x_2)$ the gluon distribution of the target. 

Nevertheless, it is important to keep in mind that this interpretation of the gauge field correlator as a gluon distribution relies on the assumption of independent scattering centers -- which is fulfilled in the context of nuclear DIS when each scattering center sits on a distinct nucleon. When dealing with a quark-gluon plasma, however, the generalization is not completely straightforward, since the medium is being formed at the same time as the initial collision is happening.

Other calculations relating the jet quenching parameter to the gluon distribution \cite{Baier:1996sk,Majumder:2007hx,CasalderreySolana:2007sw} also rely on similar assumptions, in particular on the fact that the very small-$x$ region of the target is not probed by the propagating parton.

\subsection{From single to multiple scatterings}

In order to go beyond the single scattering case it is necessary to allow multiple insertions of the background field. For the specific case of a particle propagating through the medium with only elastic collisions, it is easy to see that the resummation of the scatterings yields an exponentiation when written in coordinate space, i.e., the amplitude is given by a Wilson line. By squaring the amplitude and performing the averaging over the gauge field fluctuations of the target by using Eq.~(\ref{average}), one obtains  the probability for acquiring some  transverse momentum $\q$ after propagation through a distance $L$ in the medium 
\begin{equation}\label{transprob}
{\cal P}(\q,L)=\int d^2\r\;e^{-i\q\cdot\r}\exp\left[-\frac{N_cn}{2}L\sigma(\r)\right],
\end{equation}
where $\sigma(\r)$, the so-called dipole cross section\footnote{The exponential factor in Eq.~(\ref{transprob}) can be interpreted as the $S$-matrix for the scattering of a dipole on the medium \cite{Mueller:2012bn}.}, is related to the quantity $\gamma(\r)$ introduced in Eq.~(\ref{defgamma}) by  $\sigma(\r)=2g^2[\gamma(0)-\gamma(\r)]$. The two terms in the definition of $\sigma$ can be interpreted as real and virtual contributions in the interaction with the medium, which can be easily seen when considering the  momentum space expression,
\begin{equation}\label{sigmaq}
\frac{N_cn}{2}\sigma(\q)=-V(\q)+(2\pi)^2\delta^{(2)}(\q)\int_\l V(\l),
\end{equation}
where the second term ensures unitarity, and $\sigma(\q)=\int d^2\r\,e^{i\q\cdot\r}\sigma(\r)$. Given that this dipole cross-section satisfies $\sigma(0)=0$ and $\left.\partial_\r^i\sigma(\r)\right|_{\r=0}=0$, it is easy to see that
\begin{equation}
\int_{\q}\q^2{\cal P}(\q,L)=-\frac{N_cn}{2}L\int_\q \q^2\sigma(\q).
\end{equation}
 Here we land in the same place as in the discussion at the beginning of this section, i.e., Eq.~(\ref{avpperp}). Note that only the term linear in $\sigma$, and furthermore only the real contribution $V(\q)$ in Eq.~(\ref{sigmaq}),  contributes to $\hat q$.

It is clear from the analysis above that $\hat q$ also determines the small $\r$ behavior of the dipole cross-section $\sigma(\r)$. More specifically, the dipole cross-section is often approximated as
\begin{equation}\label{hoapprox}
\frac{N_cn}{2}\sigma(\r)\approx\frac{1}{4}\hat q(1/\r^2)\,\r^2.
\end{equation}
In the so-called harmonic approximation, often used in the context of multiple scatterings, the scale dependence of $\hat q$ is ignored, in which case the probability for acquiring a fixed transverse momentum, $\cal P$ in Eq.~(\ref{transprob}),  becomes a Gaussian:
\beq
{\cal P}(\q,L)\simeq \frac{4\pi}{\hat q L}\rme^{-\frac{\q^2}{\hat q L}}.
\eeq
 The probability distribution ${\cal P}(\q,L)$ obeys then a simple diffusion equation in which $\hat q$ plays the role of a diffusion constant. The diffusion picture is valid in the regime dominated by multiple scattering, in which a large transverse momentum is acquired by the addition of many small momentum transfers, typically when $\q^2\lesssim \hat q L$. Larger transverse momenta can be achieved through rare, single hard scattering. In the regime where $\q^2\gg \hat q L$, single collisions dominate the distribution ${\cal P}$ which is then simply given by the term linear in $\sigma$ in Eq.~(\ref{transprob}), that is (cp. Eq.~(\ref{Vdeq}))
\beq
{\cal P}(\q,L)\simeq \frac{g^4 N_c}{\q^4}n L.
\eeq

The presentation in this section is consistent with that in Ref.~\cite{Majumder:2007hx} where a similar analysis is carried out in momentum space, rather than in coordinate space as done here.

\section{Radiative corrections to transverse momentum broadening in the dense case}\label{radiativecorrection}

All the considerations in the previous section involve only collisions of a hard parton with the medium constituents, leaving aside the effects of radiation. These effects contribute to corrections to the jet quenching parameter, yielding in particular double logarithmically enhanced contributions which can be resummed for the region of phase space where the logarithms are large. Such contributions will be the subject of this section. (For other types of corrections to the interaction of a hard parton with a medium, see e.g. \cite{Ghiglieri:2015ala} and references therein.) We focus here on  contributions to $\hat q$ that come from radiative processes where an additional  gluon is emitted but is not measured. Such calculations have already been performed in a multiple scattering framework \cite{Liou:2013qya,Blaizot:2013vha,Blaizot:2014bha}. We shall  briefly review these calculations before showing how to generalize their result in such a way that regions of phase space where the single scattering mechanism is dominant can be included. We will follow the approach and notation of \cite{Blaizot:2014bha}.

\subsection{Review of calculation}

\begin{figure}
\centering
\includegraphics[width=\columnwidth]{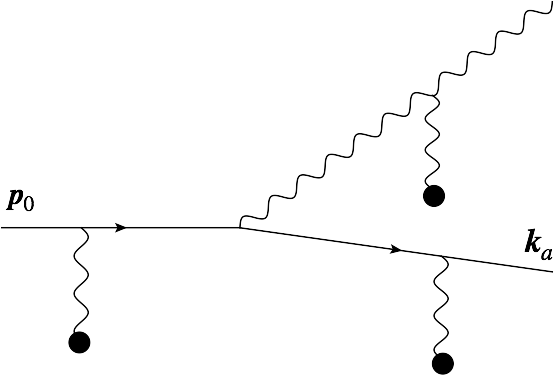}
\caption{Soft gluon emission contributing to transverse momentum broadening of a hard parton (represented by straight lines). The wavy lines represent gluons. Those that terminate on a dark blob represent collisions with the medium constituents.}
\label{figemission}
\end{figure}

\begin{figure}
\centering
\includegraphics[width=\columnwidth]{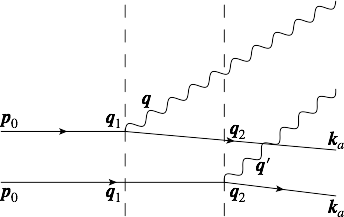}
\caption{Emission in the amplitude and conjugate amplitude with different emission times (indicated by the vertical dashed lines). Important momenta are indicated. Note in particular that $\l=\q_2-\q_1$. }
\label{figemissionsq}
\end{figure}

The basic process is the emission of a gluon in the medium, as depicted in Fig. \ref{figemission}. As stated above, the emitted gluon is not measured and therefore we do not keep track of its transverse momentum. The diagram indicates places where interactions with the medium can occur, but the more general case must be taken into account when all lines represent in-medium propagators, that is, where an arbitrary number of interactions are included. In order to exploit the fact that the medium average involves instantaneous correlations (cf. Eq.~(\ref{average})), it is convenient to draw on top of each other the process in the amplitude and the conjugate amplitude, as in Fig.~\ref{figemissionsq}, with all lines representing in-medium propagators. Scatterings only transfer transverse momenta, thereby keeping the energy of the particles intact while traversing the medium. The propagation of partons in this multiple scattering approach is then best described in a mixed representation where propagators ${\cal G}_\omega(t_1,t_0)$ depend explicitly on the time ($+$ -component in coordinate space, denoted here by $t$) and energy ($-$ -component in momentum space, denoted here by $\omega$).  

\begin{figure}
\centering
\includegraphics[width=\columnwidth]{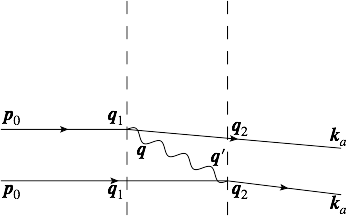}
\caption{Emission in the amplitude and conjugate amplitude after the transverse momentum of the emitted gluon has been integrated out. The vertical dashed lines (denoting the emission times in the amplitude and the complex conjugate amplitude) delineate the region where $\tilde S^{(3)}$ is defined.}
\label{figemissionint}
\end{figure}

Once the gluon is fully emitted its interactions with the medium only affect its final transverse momentum and do not affect the transverse momentum it took away from the parent particle. Therefore the propagators for this gluon after emission are not relevant for the calculation and the effective diagram to be calculated is the one in Fig.~\ref{figemissionint}. The locality of the medium average allows us to consider the three regions in the diagram separately and therefore deal with averages of two or three propagators at a time. It is clear that the average of two propagators gives precisely the momentum broadening probability of Eq.~(\ref{transprob}). The average of three propagators, to be referred to as $\tilde S^{(3)}$, is defined precisely in the Appendix.

For this calculation, the emitted gluon is taken as much softer than the parent parton but hard with respect to the medium constituents. The radiative correction to the broadening probability can be written as (details in the Appendix):

\begin{align}
&\Delta{\cal P}(\k_a,L|\p_0,0)=\;2\alpha_sN_c\text{Re}\int\frac{d\omega}{\omega^3}\int_{0}^{L}dt_2\int_{0}^{t_2}dt_1\int_{\q_1\q_2}\nn
&\quad\times{\cal P}(\k_a-\q_2,L-t_2)\nn
&\quad\times\int_{\q\q'}(\q\cdot\q')\left[\tilde S^{(3)}(\q,\q',\l+\q';t_2,t_1)-\tilde S^{(3)}(\q,\q',\l;t_2,t_1)\right]\nn
&\quad\times{\cal P}(\q_1-\p_0,t_1).\label{deltaptot}
\end{align}

In this expression, $\q$ and $\q'$ denote the momenta of the radiated gluon in the amplitude and the complex conjugate amplitude, respectively, and $\l=\q_2-\q_1$. The ${\cal P}$ factors denote the leading order momentum distributions. Note that in order to get this formula, we have retained in the splitting function only the leading soft collinear contributions, which are indeed responsible for the logarithmic enhancement\footnote{In a recent paper, Zakharov \cite{Zakharov:2018rst}   pointed out that the terms left out in this approximation may actually not be negligible  for values of the parameters that are relevant for RHIC or LHC. }.

As it was shown in the previous section, the jet quenching parameter can be obtained from the probability to acquire a fixed amount of momentum through calculating the average transverse momentum transfer (with an upper momentum cutoff). We then estimate the  correction to $\hat q$ corresponding to the radiative correction as
\begin{align}
\left\langle\Delta p_\perp^2\right\rangle=&\;\int_{\k_a}(\k_a-\p_0)^2\Delta{\cal P}(\k_a,L|\p_0,0)\nn
=&\;2\alpha_sN_c\text{Re}\int\frac{d\omega}{\omega^3}\int_{0}^{L}dt_2\int_{0}^{t_2}dt_1\nn
&\times\int_{\q\q'\l}\left[(\l-\q')^2-\l^2\right](\q\cdot\q')\tilde S^{(3)}(\q,\q',\l;t_2,t_1).\label{pt2}
\end{align}
Here it is important to note that the $\cal P$ factors appearing on the right hand side of Eq. (\ref{deltaptot}) do not play any role in the final result  \cite{Blaizot:2014bha}. This  facilitates the interpretation of the radiative corrections as a quasi local effect that can be absorbed into the jet quenching parameter $\hat q$.

At this point we use the harmonic approximation (cf. Eq.~(\ref{hoapprox})) to continue our calculation. This is consistent with the leading logarithmic accuracy we are aiming for as long as we restore the scale dependence of $\hat q$ at the end of the calculation. The 3-point correlator $\tilde S^{(3)}$ can then be calculated explicitly and the main contributions to the radiative correction can be identified. We have:
\begin{widetext}
\begin{align}
\tilde S^{(3)}(\q,\q',\l;\tau)&=\frac{16\pi}{3\hat q\tau}\exp\left\{-\frac{4[\l-(\q-\q')/2]^2}{3\hat q\tau}\right\}\frac{2\pi i}{\omega\Omega\sinh(\Omega\tau)}\exp\left\{-i\frac{(\q+\q')^2}{4\omega\Omega\coth(\Omega\tau/2)}-i\frac{(\q-\q')^2}{4\omega\Omega\tanh(\Omega\tau/2)}\right\},\label{S3ha}
\end{align}
\end{widetext}
with $\tau=t_2-t_1$ and $\Omega=\frac{1+i}{2}\sqrt{\frac{\hat q}{\omega}}$. It is clear from its functional form that this 3-point correlator is exponentially suppressed for $\tau>1/|\Omega|$: this suppression is of course related to the threshold of the multiple scattering regime and the LPM effect. Given that the main contribution to this correction comes from the small-$\omega$ region, one can safely extend the integration region for $\tau$ up to infinity without changing the result. In that case, the remaining time integration in (\ref{pt2}) gives a factor of the total length of the medium.

The radiative correction to this quantity being proportional to the length of the medium strengthens the case for considering this correction as a local effect. We have relied heavily on the soft multiple scatterings assumptions to get to this point. Without the exponential suppression argued above, which limits the formation time of the soft radiated gluon, it would not be possible to disentangle this contribution from edge effects due to emissions close to the end of the medium, and the total length would play a more important role in the result.

We can now plug this back into (\ref{pt2}) and easily perform the Gaussian integrals to get
\begin{align}
\left\langle\Delta p_\perp^2\right\rangle&=\frac{\alpha_sN_cL}{\pi}2\text{Re}\int d\omega\int d\tau\;\frac{i\Omega^3}{\sinh(\Omega\tau)}\left[1+\frac{4}{\sinh^2(\Omega\tau)}\right]\nn
&\simeq \frac{\alpha_sN_cL}{\pi}2\text{Re}\int d\omega\int d\tau\;\frac{i\Omega^3}{\Omega\tau}\nn
&\simeq\bar\alpha L\int \frac{d\omega}{\omega}\int \frac{d\tau}{\tau}\hat q\; ,\label{msdoblog}
\end{align}
where we have set $\bar\alpha\equiv \alpha N_c/\pi$, and the second line is obtained after expanding the integrand for small $\tau$ to pick up the logarithmic divergence.

\subsection{Double logarithmic phase space region}

\begin{figure*}
\centering
	\begin{subfigure}{0.47\textwidth}
	\includegraphics[width=\textwidth]{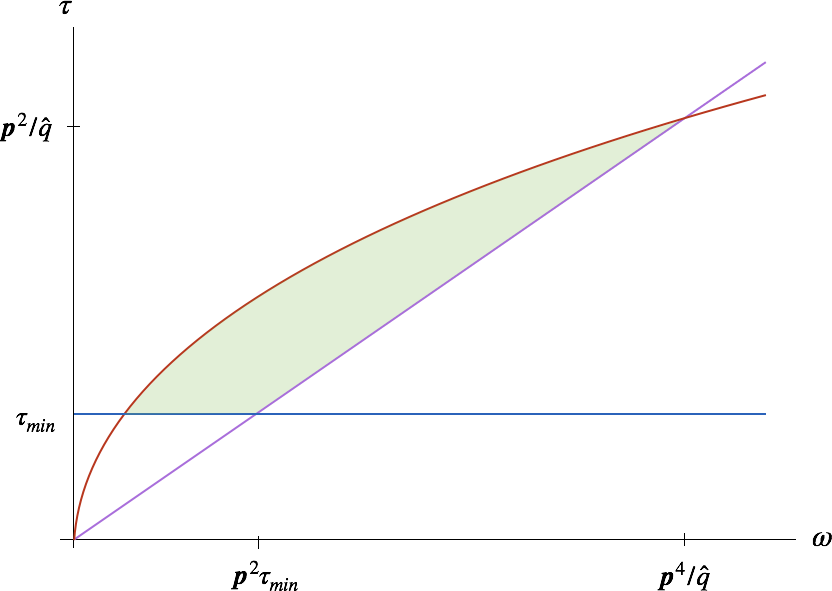}
	\caption{$(\omega,\tau)$-plane.}
	\label{figomegatau}
	\end{subfigure}
	\hspace{0.5cm}
	\begin{subfigure}{0.4\textwidth}
	\includegraphics[width=8cm]{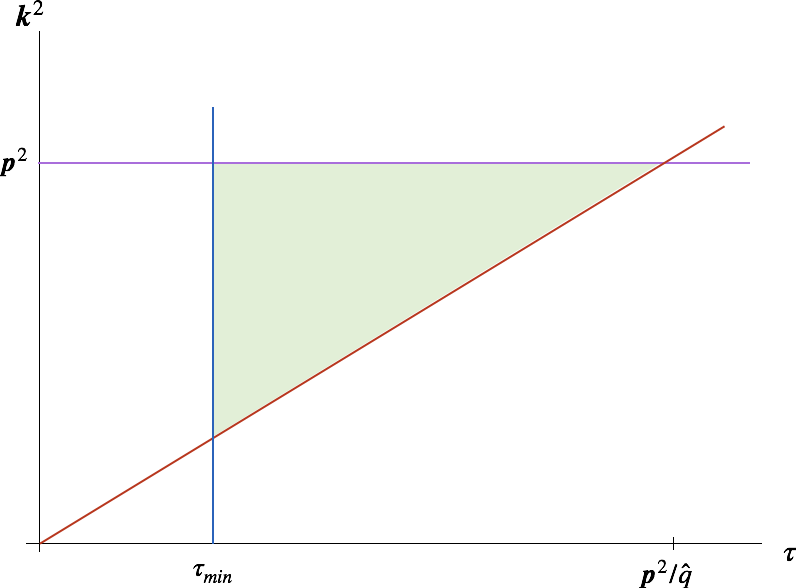}
	\caption{$(\tau,\k^2)$-plane.}
	\label{figtauksq}
	\end{subfigure}
	\caption{(Color online.) Phase space for double logarithmic contribution. The multiple scattering corresponds to the line $\tau\lesssim\sqrt{\omega/\hat q}$ (left) or $\tau\lesssim k_\perp^2/\hat q$ (right).  The formation time is given by $\tau=\omega/k_\perp^2$, and cannot be smaller, for a given $\omega$ than $\omega/\p^2$. This limit corresponds to the line $\omega=p^2\tau$ (left) and $\k^2=\p^2$ (right). }
\end{figure*}

It is clear from Eq.~(\ref{msdoblog}) that there is a double logarithmic contribution to the radiative correction and care must be taken to determine the boundaries of the relevant phase space region. In going from the first to the second lines of Eq.~(\ref{msdoblog}) it has been recognized that the main contribution to the integration comes from the region where $|\Omega\tau|$ is small. More precisely, the relevant region is limited by the condition $\tau\lesssim \sqrt{\omega/\hat q}$ which is the typical branching time for the  emission of a gluon with energy $\omega$. We may rewrite this condition as $\hat q \tau \lesssim k_\perp^2$, with $k_\perp^2=\omega/\tau$,  that is the transverse momentum acquired by the emitted gluon during its formation is less than its transverse momentum.  It has been noted \cite{Liou:2013qya,Blaizot:2013vha} that such a region corresponds to the single scattering regime, where the unmeasured gluon has a very short formation time and can scatter only once with the medium before being fully formed. The fact that the calculation is being performed in a multiple scattering setup enters as a boundary for such a region. Without this boundary, and with the approximations we have considered so far, there would not be an obvious cutoff to the logarithmic divergence.

We will postpone until the next section the question of what   happens  in the dilute case, where the probability for multiple scatterings is negligible and therefore the boundary for the double logarithmic region just mentioned is not present. First, it is important to understand the boundaries of the region of phase space contributing to the double logarithm.

These boundaries were thoroughly discussed in \cite{Liou:2013qya}  in terms of the variables $(\omega,\tau)$, and the corresponding phase space is illustrated in Fig.~\ref{figomegatau}. We agree with that discussion in general terms, but in order to be able to make the comparison with the single scattering case we have to make a subtle change to the variables involved. In \cite{Liou:2013qya}, it is assumed that the measured transverse momentum of the hard particle is of order $\hat q L$ and therefore the integration limits depend explicitly on $L$. In this paper we prefer to take the measured transverse momentum of the hard particle, $\p^2$, as an independent variable and use it as the upper limit on the integration on the transverse momentum of the soft gluon. This upper limit plays precisely the same role as the scale dependence in the leading order calculation of $\hat q$. Our choice of not equating this scale to $\hat qL$ is due to the fact that in the more general case the length $L$ sets another scale which can also introduce large logarithms. This change is important, since it emphasizes the fact that in the multiple scattering case the radiative correction is not directly sensitive to the full length of the medium and the soft gluons decorrelate from the parent parton long before the edge of the medium. When the focus is purely in particles produced via a multiple scattering mechanism, there is not much error introduced in assuming that all the transverse momenta involved in the process are of the order of $\hat qL$, but since we are interested in expanding this framework to include single scattering effects it is necessary to distinguish these two different scales and their respective effect in the jet quenching parameter.

In brief, the boundaries of the double logarithmic region in this set of coordinates are given by restrictions over the formation time of the radiated gluon. On the one hand it cannot be too large to prevent the gluon to scatter more than once, and on the other hand it cannot be too short so as to allow the gluon to see the medium or to be resolved from its parent.

In \cite{Blaizot:2014bha}, it was noted that it was more convenient to switch to the set of variables $(\tau,\k^2)$ with $\tau=\omega/\k^2$, see Fig. \ref{figtauksq}. In this set of variables, the boundaries of the double logarithmic region have a straightforward interpretation: a lower bound on the interaction time with the medium, the multiple scattering condition which does not allow long formation times, and an upper cutoff in the momentum integration which plays the role of the scale dependence of $\hat q$ and the resolution scale in transverse space. The lower bound in the formation time plays an important role, since otherwise the medium would not be able to resolve the fluctuation. Lower formation times require larger fractions of energy being carried by the gluons colliding with the probe, which is outside the scope of our approximations.

This change of variables makes the evaluation of the integral easier to understand and allows us to connect with the discussion of the average transverse momentum transfer at leading order where the scale dependence of $\hat q$ was introduced. Despite the fact that the bulk of the calculation was performed in the harmonic approximation where $\hat q$ is regarded as a constant, we can safely consider a $k_\perp$-dependent $\hat q$ for the evaluation of (\ref{msdoblog}) since we are only interested in the leading logarithmic behavior.

The correction to the jet quenching parameter then takes the following form:
\begin{equation}\label{msdoblogtauk}
\Delta\hat q=\bar\alpha\int_{\tau_{\text{min}}}^{\p^2/\hat q}\frac{d\tau}{\tau}\int_{\hat q\tau}^{\p^2}\frac{d\k^2}{\k^2}\hat q(\k^2).
\end{equation}
For the case where the scale dependence of $\hat q$ is ignored, one gets
\begin{equation}\label{msdoblogtauk2}
\Delta\hat q=\frac{\bar\alpha}{2}\,\hat q \ln^2\frac{\p^2}{\hat q\tau_{\text{min}}}.
\end{equation}

Before moving on to the calculation in the dilute case, it is worth noting that this approach has been extended \cite{Mueller:2016xoc,Iancu:2014sha}, still in the multiple scattering framework, to the case where $\p^2>\hat q L$ in an \emph{ad hoc} way by requiring that the formation time of the unmeasured gluon  be shorter than the length of the medium. Even though this is a sensible assumption and the results are indeed correct, it is clear that in that region of phase space some of the approximations made here are not valid, and therefore the derivation must be revisited in the framework of an opacity expansion calculation where the finite length effects must be properly taken into account.

\begin{figure}
\centering
\includegraphics[width=\columnwidth]{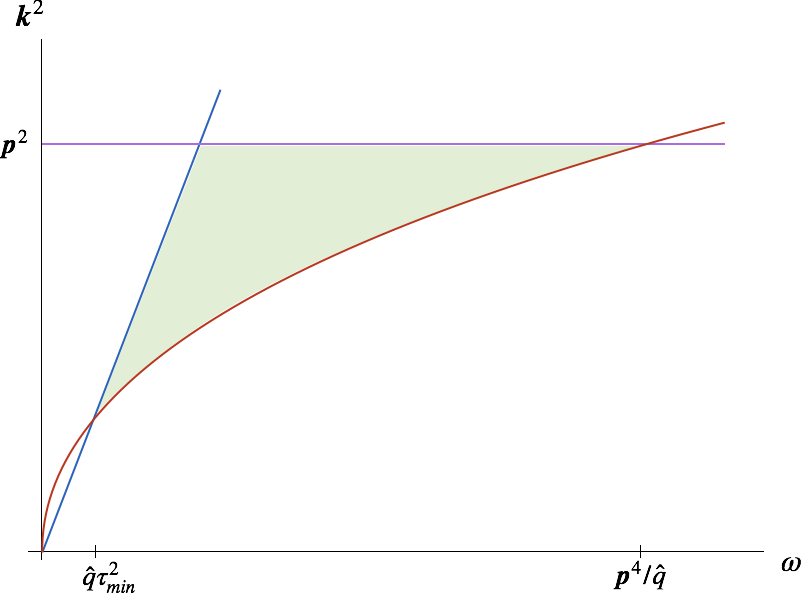}
\caption{(Color online.) Phase space for double logarithmic contribution in the $(\omega,\k^2)$-plane.}
\label{figomegaksq}
\end{figure}

In the same line as previous discussions in this section, it will be seen in the following section that for the dilute case the variable $\tau$ is not as convenient since the structure of the time integration is very different. The more natural set of variables for that case is then $(\omega,\k^2)$. In order to be able to make a clear comparison between the two different cases we show in Fig. \ref{figomegaksq} the double logarithmic region for the dense case in that particular set of variables.

\section{Radiative corrections to transverse momentum broadening in the dilute case}\label{dilutecase}

Since it was already seen in the previous section that the phase space region for single scattering plays an essential role in the determination of the double logarithmic enhancement, it will be useful to follow the same set up as for the multiple scattering case,  while keeping close attention to the fundamental differences of the two derivations.

\subsection{Calculation of radiative correction}

First, we note that, in principle, the single scattering result can be obtained by taking as a starting point the multiple scattering framework, and expanding in terms of the number of scatterings. This approach will work as long as the expansion is performed before we consider approximations which are valid only for the soft multiple scattering case, such as the harmonic approximation.

In particular, there are two features of the single scattering regime that the multiple scattering formalism usually misses. First, large transfers of transverse momentum, the so-called ``perturbative tail'', are, in the multiple scattering setup,  exponentially suppressed rather than following the power law behavior predicted by pQCD. Second, the formation time of radiated gluons  is no longer governed by the scatterings with the medium, and fluctuations can be long-lived and see the full length of the medium.

This second feature has important consequences in our analysis and the interpretation of the results, since it breaks the picture of semi-local corrections due to short-lived fluctuations which can occur with equal probability anywhere inside the medium (giving therefore a contribution that is proportional to the length of the medium). Accordingly, the concept of a transition probability for a fixed length is not useful anymore since it becomes important to consider the effects of emissions even after the particle has exited the medium. As was argued before, the relevant quantity to calculate is the average transverse momentum transfer, as was done in Eq.~(\ref{pt2}). In order to take advantage of the framework set for the multiple scattering case, we will only consider the case where we let the emission times run up to infinity, even though the interaction with the medium is still confined to a finite region, keeping in mind that the medium propagators and their correlators become their vacuum versions when considered outside the medium.

Our starting point for the opacity expansion is therefore the last line of Eq.~(\ref{pt2}) with the integration over $t_2$ allowed to run up to infinity, with a regulator $e^{-\epsilon(t_1+t_2)}$ turning off the interactions adiabatically. In order to find the single scattering contribution we expand the correlator $\tilde S^{(3)}$, keeping terms up to first order in the dipole cross section,
\begin{widetext}
\begin{align}
\tilde S^{(3)}(\q,\q',\l;t_2,t_1)&\approx(2\pi)^4\delta^{(2)}(\l)\delta^{(2)}(\q-\q')e^{-i\frac{\q^2}{2\omega}(t_2-t_1)}-\frac{N_c}{4}\int_{t_1}^{t_2}dt\;n\,e^{-i\frac{\q^{\prime 2}}{2\omega}(t_2-t)-i\frac{\q^2}{2\omega}(t-t_1)}\nn
&\qquad\qquad\qquad\qquad\times (2\pi)^2\left[\delta^{(2)}(\l)\sigma(\q-\q')+\delta^{(2)}(\q-\q'+\l)\sigma(\l)+\delta^{(2)}(\q-\q')\sigma(\l)\right]. \label{expS3}
\end{align}
\end{widetext}

After plugging this expansion in Eq.~(\ref{pt2}) we are left with a triple time integral, the time variables corresponding to the emission in the amplitude, the emission in the conjugate amplitude, and the interaction with the medium. In the way the calculation is set up, the latter integration is the inner most time integral. In order to properly deal with the regularization of emissions at infinity, it is better to change the integration order and make the interaction time the outer most integration. The time integrals that are needed in the rest of the calculation take the form:
\begin{align}
&\text{Re}\int_{0}^{L}dt\int_{0}^t dt_1\int_{t}^\infty dt_2 e^{-i\frac{\q^{\prime2}}{2\omega}(t_2-t)-i\frac{\q^2}{2\omega}(t-t_1)}e^{-\epsilon t_2}\nn
&\qquad=-\frac{4\omega^2}{\q^2\q^{\prime2}}\left(L-\frac{2\omega}{\q^{\prime2}}\sin\left[\frac{\q^{\prime2}}{2\omega}L\right]\right).
\end{align}

Using the equations above it is straightforward to calculate the radiative correction to the average momentum transfer in the single scattering case. One obtains (see Appendix for details)
\begin{align}
\left\langle\Delta p_\perp^2\right\rangle&=2\alpha_sN_c^2n\int\frac{d\omega}{\omega}\int_{\q'\q}\sigma(\q'-\q)\frac{2(\q'\cdot\q)^2}{\q^{\prime 2}\q^2}\nn
&\quad\times\left\{L-\frac{2\omega}{\q^{\prime 2}}\sin\left[\frac{\q^{\prime 2}}{2\omega}L\right]\right\}.\label{Deltaq2}
\end{align}

A couple of comments about the derivation above are due: first, we have purposefully left the above formula without integration limits. The next subsection is devoted to specify the phase space region in the physically relevant case where we expect in particular a double logarithmic enhancement. Second, the simplicity of the derivation lies in the fact that we chose to make the opacity expansion for the average momentum transfer, where the factors accounting for transverse momentum broadening have already been integrated out. Making the expansion at any other previous stage of the calculation would have made it lengthier since multiple factors would have to be expanded. In the Appendix such a calculation is sketched, where it is noted that multiple cancelations occur when the momentum average is taken.

\subsection{Phase space region for double logarithmic contribution}

The result from the previous section allows us to properly take into account the  effects of a medium of finite size on the radiative correction. First of all, it is easy to see that the first term, proportional to the medium length, will show the same behavior as the double logarithm found in the multiple scattering case. In fact, it coincides with the analysis performed in the Appendix of Ref.~\cite{Blaizot:2013vha} where it is recognized that the double logarithmic radiative correction corresponds to the single scattering region. In summary,
\begin{align}
\int\frac{d\omega}{\omega}\int_{\k'\k}\sigma(\k'-\k)&\frac{2(\k'\cdot\k)^2}{\k^{\prime 2}\k^2}\nn
&=\int\frac{d\omega}{\omega}\int_{\l\k}\sigma(\l)\frac{2((\k+\l)\cdot\k)^2}{(\k+\l)^2\k^2}\nn
&=2\int\frac{d\omega}{\omega}\int_{\l\k}\sigma(\l)\frac{(\l\cdot\k)^2-\l^2\k^2}{(\k+\l)^2\k^2}\nn
&\sim-\int\frac{d\omega}{\omega}\int_\k\frac{1}{\k^2}\int_\l\l^2\sigma(\l).\label{ssdoubllog}
\end{align}
In going from the second to the third line above, we have recognized that the logarithmic enhancement is present for $\k\gg\l$. This sets the upper boundary for the $\l$-integration, which will set the scale for $\hat q$ as in Eq.~(\ref{hatqdef}).

Plugging the result of Eq.~(\ref{ssdoubllog}) back into Eq.~(\ref{Deltaq2}), one gets the total contribution in the region where the sine function oscillates rapidly and therefore does not contribute. On the other hand, the oscillatory term induced by the finite size effects will provide the missing boundary to the double logarithmic region. If the argument of the sine function is small, one can perform a power expansion, where the first non-zero contribution will be accompanied by an extra power of momentum which kills the logarithmic contribution. In the relevant set of variables ($\omega,\k^2$), the new boundary is then given by the condition
\begin{equation}\label{ssbound}
\frac{\k^2}{2\omega}L>1.
\end{equation}

\begin{figure}
\centering
\includegraphics[width=\columnwidth]{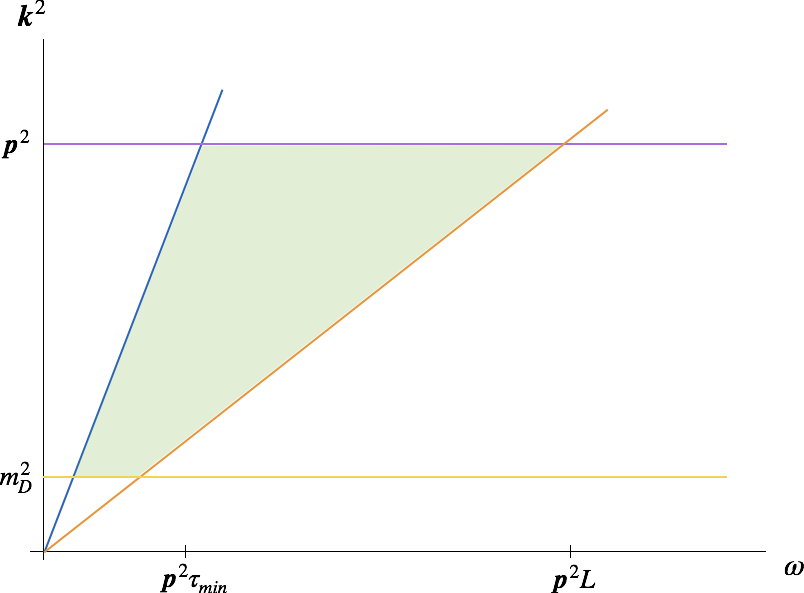}
\caption{(Color online.) Phase space for double logarithmic contribution in single scattering case}
\label{figomegaksqss}
\end{figure}

This new boundary effectively replaces the one provided by the multiple scattering condition for the case in hand. The corresponding region is displayed in Fig. \ref{figomegaksqss}. It is easy to see that this condition is equivalent to the requirement that the formation time of the gluon is shorter than the length of the medium, which is the condition used in \cite{Mueller:2016xoc,Iancu:2014sha} in a multiple scattering setting to go beyond the region $\k^2\sim\hat qL$. It is important to emphasize that we have here a clear derivation for this boundary for the single scattering case, which is the relevant physics for this regime of large transverse momentum transfer.

The boundary given by Eq.~(\ref{ssbound}) does not completely replace the multiple scattering boundary. It is better seen as an upper boundary on the energy of the emitted gluon, and it allows arbitrarily low values for the transverse momentum. In that case, the natural infrared cut off for the transverse momentum integration is given by the Debye mass (setting the lower bound in Fig. \ref{figomegaksqss}), signaling the scale at which screening sets in.

Combining this with Eqs.~(\ref{Deltaq2}), (\ref{ssdoubllog}), and (\ref{ssbound}), it is easy to find the leading logarithmic contribution to the radiative correction to the transverse momentum broadening,
\begin{align}
\left\langle\Delta p_\perp^2\right\rangle&=\bar\alpha L\int_{m_D^2}^{\p^2}\frac{d\k^2}{\k^2}\int_{\k^2\tau_{\text{min}}}^{\k^2L}\frac{d\omega}{\omega}\hat q(\k^2).
\end{align}
As in the multiple scattering case, $\p^2$ enters as the upper limit of integration for the transverse momentum, therefore setting the scale at which the correction is evaluated. Nevertheless, there is a qualitative difference with the result in the multiple scattering case: the length of the medium enters explicitly in the limits of integrations independently of $\p^2$, therefore setting an additional scale for large logarithms. Here it is instructive to go back to the set of variables $(\tau,\k^2)$. Even though the calculation is not naturally set in those variables, it allows us to make a direct comparison to Eq.~(\ref{msdoblogtauk}).
\begin{align}
\left\langle\Delta p_\perp^2\right\rangle&=\bar\alpha L\int_{\tau_{\text{min}}}^{L}\frac{d\tau}{\tau}\int_{m_D^2}^{\p^2}\frac{d\k^2}{\k^2}\hat q(\k^2).
\end{align}

\begin{figure}
\centering
\includegraphics[width=\columnwidth]{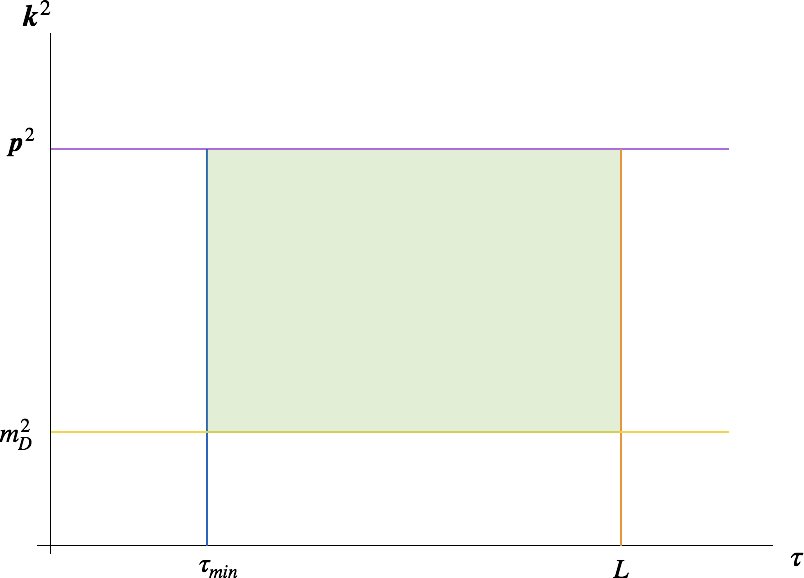}
\caption{(Color online.) Phase space for double logarithmic contribution in single scattering case in $(\tau,\k^2)$-plane.}
\label{figtauksqss}
\end{figure}

Here it becomes clear that the two integrals decouple, each one giving one of the logarithms, one controlled by the momentum scale $\p^2$ and the other by the length of the medium, as can also be seen when we sketch the region of integration in this set of variables as in Fig. \ref{figtauksqss}. The structure is reminiscent of the double logarithmic approximation of DGLAP at small-$x$ \cite{Dokshitzer:1991wu} but the main difference relies in the ordering variables. The ordering in formation time changes considerably the dynamics of the process and sets the medium length as the limit for the phase space.

For the case where the scale dependence of $\hat q$ is ignored inside the integral, the result is then
\begin{align}
\frac{\left\langle\Delta p_\perp^2\right\rangle}{L}&=\bar\alpha\hat q\ln\frac{\p^2}{m_D^2}\ln\frac{L}{\tau_{\text{min}}}.\label{ssqhatconst}
\end{align}
This result can be compared with the corresponding result for the multiple scattering case in Eq. (\ref{msdoblogtauk2}) where the two logarithms are not independent.

\section{Finite size effects vs multiple scattering}\label{finitesize}

From the previous section, it is clear that there is room for a double logarithmic contribution from radiative corrections to transverse momentum broadening in the single scattering case. Nevertheless, there is a conceptual problem with the way the region is set up. Since the relevant variable determining the size of the phase space region is the length of the medium, the double logarithmic contribution is enhanced only for a large medium, where the use of the single scattering formalism is not justified. In other words, to have a large logarithmic contribution one needs to have a large medium, but in a large medium the probability to have more than one scattering is large. This may be one of the reasons why the current formalisms used to describe jet quenching phenomena based on an opacity expansion do not include such double logarithmic contributions.

Now, taking into account that the two approximations studied in this paper are somewhat complementary and can be applied in different regions of phase space, it becomes important to understand how the multiple soft scattering and the single hard scattering approaches can be combined to obtain the most general region of phase space with a double logarithmic enhancement. When the corresponding phase space regions are superimposed, we see that the multiple scattering case is more restrictive at low transverse momentum while the single scattering case is more restrictive at larger momentum. This might seem counterintuitive at first, since in principle through multiple scatterings the gluon could accumulate more transverse momentum, but what actually occurs is that in order to have a large momentum transfer one always needs (at least) one hard kick from the medium which dominates over multiple soft ones.

\begin{figure}
\centering
\includegraphics[width=\columnwidth]{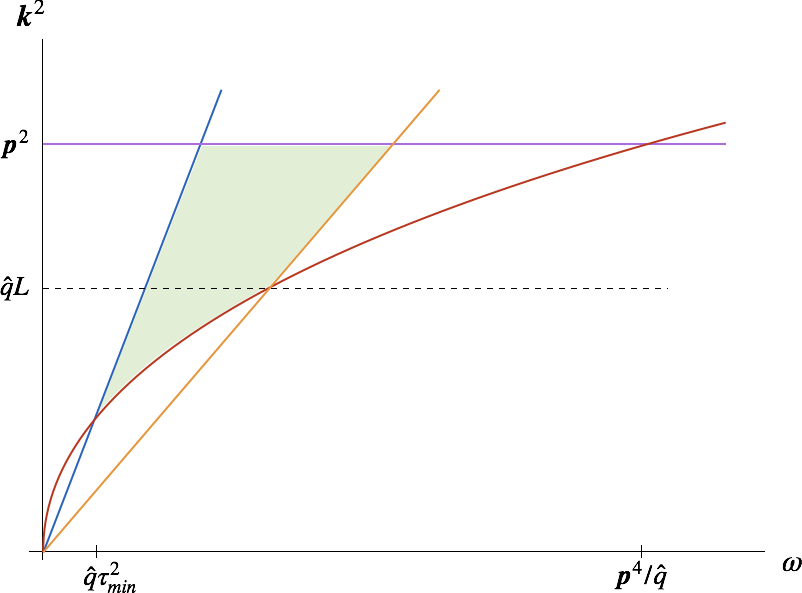}
\caption{(Color online.) Phase space when single scattering dominates over multiple scattering}
\label{figssvsms}
\end{figure}

It is easy to see that the corresponding boundaries for the two phase space regions intersect at $\k^2=\hat qL$. If the momentum scale $\p^2$ is smaller than this value then the multiple scattering regime dominates. The medium is too long for the gluon to only scatter once. But if the momentum scale is larger, then there is a region of phase space that is only accesible through a single hard scattering. This division is shown in Fig. \ref{figssvsms} and the corresponding double integral is (for the case where the scale dependence of $\hat q$ is neglected)

\begin{align}
&\int_{\hat q\tau_{\text{min}}}^{\hat qL}\frac{d\k^2}{\k^2}\int_{\k^2\tau_{\text{min}}}^{\k^4/\hat q}\frac{d\omega}{\omega}+\int_{\hat qL}^{\p^2}\frac{d\k^2}{\k^2}\int_{\k^2\tau_{\text{min}}}^{\k^2L}\frac{d\omega}{\omega}\nn
&=\frac{1}{2}\ln^2\frac{L}{\tau_{\text{min}}}+\ln\frac{L}{\tau_{\text{min}}}\ln\frac{\p^2}{\hat qL}.\label{mixdoublog}
\end{align}

\begin{figure}
\centering
\includegraphics[width=\columnwidth]{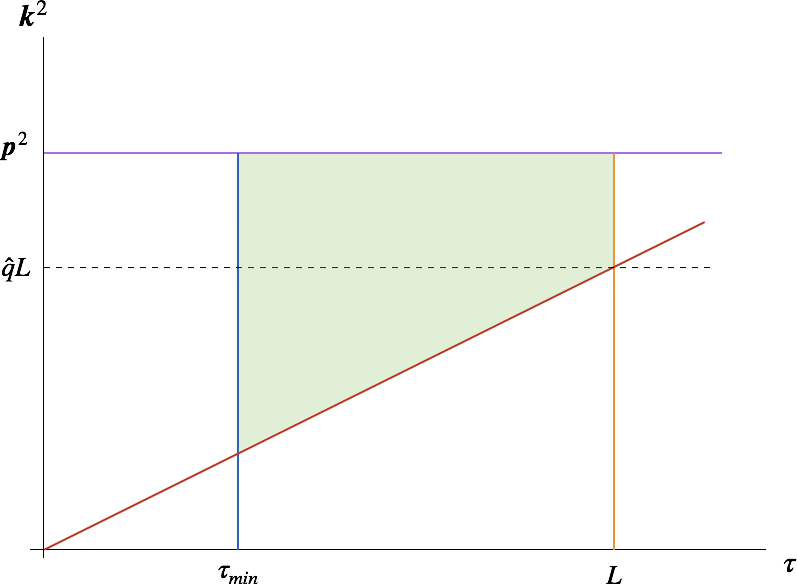}
\caption{(Color online.) Phase space when single scattering dominates over multiple scattering in the $(\tau,\k^2)$-plane.}
\label{figssvsmstauksq}
\end{figure}

As was already noted in the previous section, it is easier to understand how this region is split in the $(\tau,\k^2)$-plane as shown in Fig. \ref{figssvsmstauksq}. The first term in (\ref{mixdoublog}) corresponds to the triangle below $\hat qL$ and is reminiscent of the multiple scattering contribution (c.f. Eq. (\ref{msdoblogtauk2})), while the second term corresponds to the rectangle above and represents the contribution from the region where single scattering dominates (c.f. Eq. (\ref{ssqhatconst})). The relative factor of $1/2$ explicitly reflects the difference in the shapes of the two regions.

It is the second term in (\ref{mixdoublog}) the one which shows the effect of the single scattering case on top of the already known multiple scattering contribution. It clearly shows how, for a fixed $\p^2$, changing the length of the medium could enhance one of the logarithms but would necessarily suppress the other one. Moreover, the ratio of the two terms grows as the log of the length $L$, showing that for larger $L$ the multiple scattering contribution is parametrically more important. This was already known, since for large media edge effects play a secondary role and the formation time of the emitted gluons will always occur completely inside the medium.

\section{Conclusions}\label{conclusions}

In this paper, we have examined the current calculations of radiative corrections to in-medium gluon emission to leading logarithmic accuracy and have shown that it is possible to go beyond the multiple soft scattering approximation in order to properly include the region of phase space where a single hard scattering dominates. This finding shows that the radiative correction can be important in a wider range of processes, since it is not only limited to the case where the medium is either very large or very dense, and could play a role in realistic conditions.

Previous calculations, in frameworks where only one or a few scatterings where taken into account, have failed to identify similar logarithmically enhanced radiative corrections, perhaps because of the intricacies of the particular formalisms used in these calculations.  Our calculation takes a rather simple form by exploiting  the formalism widely used for multiple scatterings\footnote{This formalism is known to work well for any order in opacity, see e.g. \cite{Wiedemann:2000za}}, and taking advantage of several cancellations that occur for the leading logarithmic region.   For this   calculation to work, it is important to recognize that the main contribution to the double logarithmic region comes from very soft emissions, where the fraction of energy taken by the emitted gluon is kept only in the divergent part of the splitting function and set to zero everywhere else. The emissions responsible for the logarithmic enhancement are those with a short formation time and with a transverse momentum larger than the typical accumulated momentum due to collisions. They are therefore not very collinear. This might be the reason why other approaches, in particular those based on a higher twist expansion \cite{Kang:2014ela} that involve an expansion around the collinear direction, do not reproduce the double logarithmic enhancement.

\begin{acknowledgements}

We would like to thank Al Mueller and Yacine Mehtar-Tani for many useful discussions. FD was supported by a Marie Sklodowska-Curie Individual Fellowship of the European CommissionÕs Horizon 2020 Programme under contract number 660952 HPpQCD, Ministerio de Ciencia e Innovaci\'on of Spain under projects FPA2014-58293-C2-1-P, FPA2017-83814-P, Unidad de Excelencia Mar\'ia de Maetzu under project MDM-2016-0692, by Xunta de Galicia (Conseller'a de Educaci—n) within the Strategic Unit AGRUP2015/11, and by FEDER.

\end{acknowledgements}

\appendix*

\section{Calculational details}

In this appendix some of the technical details on the calculations presented in the main text are provided. Further details on the techniques and notation used in this paper can be found in Refs.~\cite{Blaizot:2013vha,Blaizot:2014bha}.

\subsection{Multiple scattering case}

In the multiple scattering formalism, the main building block for the calculation is the in-medium gluon propagator, that is the propagator of a high energy gluon in a fluctuating soft background field $A^-(t, \r(t))$. This is given by the following path integral
\begin{align}
&(\x_1|{\cal G}(t_1,t_0)|\x_0)\nn
&=\theta(t_1-t_0)\int{\cal D}\r\;\exp\left(\frac{i\omega}{2}\int_{t_0}^{t_1}dt\,\dot\r^2\right)U_\r(t_1,t_0),
\end{align}
where the paths $\r(t)$ satisfy $\r(t_0)=\x_0$, $\r(t_1)=\x_1$, and $U_\r$ is a Wilson line in the adjoint representation along the path $\r(t)$:
\begin{equation}
U_\r(t_1,t_0)=\text{T}\exp\left[ig\int_{t_0}^{t_1}dt\, A^-\left(t,\r(t)\right)\right].
\end{equation}
In this equation the sympbol $\rm T$ denotes path ordering, with the variable $t$ representing the light cone coordinate $x^+$ (often referred to as ``time'' in this paper).

To leading order, the probability ${\cal P}(\p_1,t_1|\p_0,t_0)$ for a hard gluon to acquire a fixed quantity of transverse momentum, $\p_1-\p_0$, as it traverses the medium during the time $t_1-t_0$, is directly related to the medium average of two gluon propagators
\begin{align}
&(2\pi)^2\delta^{(2)}(\p_0-\bar\p_0){\cal P}(\p_1,t_1|\p_0,t_0)\nn
&=\int_{\x_1\x_0\bar\x_1\bar\x_0}e^{-i\p_1\cdot(\x_1-\bar\x_1)+i\p_0\cdot\x_0-i\bar\p_0\cdot\bar\x_0}(X_1|S^{(2)}(t_1,t_0)|X_0),\label{pbroadprob}
\end{align}
with
\begin{align}
&(X_1|S^{(2)}(t_1,t_0)|X_0)\nn
&=\frac{1}{N_c^2-1}\left\langle\text{Tr}(\x_1|{\cal G}(t_1,t_0)|\x_0)(\bar\x_0|{\cal G}^\dagger(t_0,t_1)|\bar\x_1)\right\rangle,
\end{align}
where the trace runs over the color degrees of freedom, and  $\x_i, \bar \x_i$ ($i=0,1$) denote the coordinates of the hard gluon in the amplitude and the complex conjugate amplitude, respectively. 
The brakets in the equation above denote the mediurm average, i.e., the average over the fluctuating gauge field. This is calculated using Eq.~(\ref{average}) and Wick's theorem. Noticing that the multiple scattering series exponentiates, one gets Eq.~(\ref{transprob}) after a straightforward calculation.

The leading radiative correction involves the emission of a gluon which is not measured. Its evaluation requires that of the medium average of three gluon propagators:
\begin{widetext}
\beq\label{S3def}
(X_1|S^{(3)}(t_1,t_0)|X_0)=\frac{f^{a_1b_1c_1}f^{a_0b_0c_0}}{N_c(N_c^2-1)}\left\langle(\y_1|{\cal G}^{a_1a_0}(t_1,t_0)|\y_0)(\x_1|{\cal G}^{b_1b_0}(t_1,t_0)|\x_0)(\bar\x_0|{\cal G}^{\dagger c_0c_1}(t_0,t_1)|\bar\x_1)\right\rangle.
\eeq
\end{widetext}
As it is explained in \cite{Blaizot:2014bha}, this three-point function contributes ``real'' and ``virtual''  corrections to the broadening probability $\Delta{\cal P}$. The real contribution is  given by
\begin{widetext}
\begin{align}
\Delta{\cal P}_r(\k_a,L|\p_0,0)=&\;2\alpha_sN_c\text{Re}\int\frac{d\omega}{\omega^3}\int_{0}^{L}dt_2\int_{0}^{t_2}dt_1\int_{\p_1\q_1\bar\q_2\q_2}(\p_1-\q_1)\cdot(\q_2-\bar\q_2)\nn
&\times{\cal P}(\k_a-\q_2,L-t_2)\tilde S^{(3)}(\p_1-\q_1,\q_2-\bar\q_2,\bar\q_2-\q_1;t_2,t_1){\cal P}(\q_1-\p_0,t_1),\label{realterm}
\end{align}
\end{widetext}
where only the leading behavior in the limit $z\to 0$ is kept, with $z$ the fraction of longitudinal momentum taken by the emitted gluon, and $\omega$ the energy of that gluon. The $\tilde S^{(3)}$ factor in the equations above corresponds to the Fourier transform of (\ref{S3def}) after an overall transverse-momentum conserving delta-function has been removed (see \cite{Blaizot:2014bha} for details). The virtual contribution reads
\begin{widetext}
\begin{align}
\Delta{\cal P}_v(\k_a,L|\p_0,0)=&-2\alpha_sN_c\text{Re}\int\frac{d\omega}{\omega^3}\int_{0}^{L}dt_2\int_{0}^{t_2}dt_1\int_{\p_1\q_1\bar\q_2\q_2}(\p_1-\q_1)\cdot(\q_2-\bar\q_2)\nn
&\times{\cal P}(\k_a-\bar\q_2,L-t_2)\tilde S^{(3)}(\p_1-\q_1,\q_2-\bar\q_2,\bar\q_2-\q_1;t_2,t_1){\cal P}(\q_1-\p_0,t_1).\label{virtualterm}
\end{align}
\end{widetext}
These two contributions (real plus virtual) add up to yield Eq.~(\ref{deltaptot}), after a simple change of variables (see \cite{Blaizot:2014bha}).

\subsection{Single scattering case}

As indicated in the main text, the simplicity of the calculation presented for the single scattering case relies on the fact that most of the steps taken to arrive at Eq.~(\ref{pt2}) do not rely on approximations specific to the multiple soft scattering case. In particular, there are two key observations that allow us to anticipate that contributions from several different diagrams cancel out. First, diagrams where the emitted gluon interacts with the medium after it has been emitted both in the amplitude and the conjugate amplitude cancel out when its transverse momentum is integrated over. Second, the broadening of the parent parton does not contribute either to the radiative correction for soft enough emissions, since the real and virtual contribution exactly cancel.

The first of these observations can be understood from the fact that the interactions with the medium only change the final transverse momentum of the emitted gluon but do not affect the probability of being emitted nor the amount of transverse momentum it takes away from the parent particle. The other observation is harder to justify without a more explicit calculation of the transition probability, which will be sketched here.

The starting point would be the expressions (\ref{realterm}) and (\ref{virtualterm}) for the real and virtual contributions, with the upper limit of the $t_2$ integration sent to infinity, with an appropriate adiabatic cutoff. In addition to the expansion of the three point function, Eq.~(\ref{expS3}), we need also the first order expansion of the factors ${\cal P}$, given by
\begin{equation}
{\cal P}(\q,t_f-t_i)\approx(2\pi)^2\delta^{(2)}(\q)-\frac{N_c}{2}\int_{t_i}^{t_f}dt\;n\,\sigma(\q).
\end{equation}
The following time integrals are necessary when the order of integration has been changed in the same way as explained in the main text:
\begin{align}
&\text{Re}\int_{0}^{L}dt\int_t^\infty dt_1\int_{t_1}^\infty dt_2 \, e^{-i\frac{\p^2}{2\omega}(t_2-t_1)}e^{-\epsilon(t_1+t_2)}\nn
&\hspace{2cm}=\frac{2\omega^2}{\p^4}L,\\
&\text{Re}\int_{0}^{L}dt\int_{0}^t dt_1\int_{t_1}^t dt_2 \,e^{-i\frac{\p^2}{2\omega}(t_2-t_1)}\nn
&\hspace{2cm}=\frac{4\omega^2}{\p^4}\left(L-\frac{2\omega}{\p^2}\sin\left[\frac{\p^2}{2\omega}L\right]\right).
\end{align}
Plugging these various expressions into Eq.~(\ref{realterm}) and keeping the terms which are of first order in opacity (i.e. first order in the dipole cross section) we get:
\begin{widetext}
\begin{align}
\Delta{\cal P}_r(\k_a,\infty|\p_0,0)&=-2\alpha_sN_c^2n\int\frac{d\omega}{\omega}\int_{\p_1}\sigma(\p_1+\k_a-\p_0)\left\{\frac{1}{\p_1^2}L+\frac{1}{\p_1^2}\left(1-\frac{\p_1\cdot(\p_0-\k_a)}{(\p_0-\k_a)^2}\right)\left(L-\frac{2\omega}{\p_1^2}\sin\left[\frac{\p_1^2}{2\omega}L\right]\right)\right.\nn
&\quad\left.-\frac{\p_1\cdot(\p_0-\k_a)}{\p_1^2(\p_0-\k_a)^2}\left(L-\frac{2\omega}{(\p_0-\k_a)^2}\sin\left[\frac{(\p_0-\k_a)^2}{2\omega}L\right]\right)\right\}.\label{1opreal}
\end{align}

Proceeding similarly  for the virtual correction, Eq.~(\ref{virtualterm}), one gets
\begin{align}
\Delta{\cal P}_v(\k_a,\infty|\p_0,0)&=2\alpha_sN_c^2n\int\frac{d\omega}{\omega}\left\{\sigma(\k_a-\p_0)\int_{\p_1}\left[\frac{1}{\p_1^2}L+\frac{1}{\p_1^2}\left(1-\frac{\p_1\cdot(\p_1+\k_a-\p_0)}{(\p_1+\k_a-\p_0)^2}\right)\left(L-\frac{2\omega}{\p_1^2}\sin\left[\frac{\p_1^2}{2\omega}L\right]\right)\right]\right.\nn
&\left.\quad+(2\pi)^2\delta^{(2)}(\k_a-\p_0)\int_{\p_1\q_2}\sigma(\q_2-\p_1)\frac{\p_1\cdot\q_2}{\p_1^2\q_2^2}\left(L-\frac{2\omega}{\p_1^2}\sin\left[\frac{\p_1^2}{2\omega}L\right]\right)\right\}.\label{1opvirtual}
\end{align}
\end{widetext}

The real and virtual terms do not combine in any particular useful way to yield a simple expression for the broadening probability. However simplifications occur when we combine these two contributions in the calculation of the average transverse momentum transfer, i.e.
\begin{align}
&\left\langle\Delta p_\perp^2\right\rangle=\int_{\k_a}(\k_a-\p_0)^2\nn
&\quad\times\left[\Delta{\cal P}_r(\k_a,\infty|\p_0,0)+\Delta{\cal P}_v(\k_a,\infty|\p_0,0)\right].\label{1oppt2}
\end{align}
Now look at the contributions from the first term of Eq.~(\ref{1opreal}) and Eq.~(\ref{1opvirtual}) to Eq.~(\ref{1oppt2}). They look the same except for an opposite sign and a displacement of the argument of the $\sigma$ factor. Nevertheless, due to the identities $\int_{\k}\sigma(\k)=\int_{\k}\k\sigma(\k)=0$, we have 
\begin{equation}
\int_{\k_a}(\k_a-\p_0)^2\sigma(\p_1+\k_a-\p_0)=\int_{\k_a}(\k_a-\p_0)^2\sigma(\k_a-\p_0),
\end{equation}
and therefore these two terms cancel exactly. It is easy to see that these particular terms correspond to the expansion of the first $\cal P$ factors appearing in both Eq.~(\ref{realterm}) and Eq.~(\ref{virtualterm}).

Similar cancelations occur with other terms, including those corresponding to the expansion of the other $\cal P$ factor. The only remaining contributions come from the interference of diagrams where the interaction occurs before the emission with diagrams where the interaction occurs after the emission. This is equivalent to the observation made after Eq.~(\ref{pt2}) in the multiple scattering case, where it is noted that the broadening before and after the emission do not play a role in the calculation of the radiative correction.

After accounting for these cancelations, the only surviving terms are those coming from the sine term in the last line of Eq.~(\ref{1opreal}) and one of the terms in the second line of Eq.~(\ref{1opvirtual}) yielding together the result in Eq.~(\ref{Deltaq2}).


\begin{thebibliography}{10}
  

\bibitem{Liou:2013qya} 
  T.~Liou, A.~H.~Mueller and B.~Wu,
  Nucl.\ Phys.\ A {\bf 916}, 102 (2013)
  doi:10.1016/j.nuclphysa.2013.08.005
  [arXiv:1304.7677 [hep-ph]].


\bibitem{Blaizot:2013vha} 
  J.~P.~Blaizot, F.~Dominguez, E.~Iancu and Y.~Mehtar-Tani,
  JHEP {\bf 1406}, 075 (2014)
  doi:10.1007/JHEP06(2014)075
  [arXiv:1311.5823 [hep-ph]].


\bibitem{Blaizot:2014bha} 
  J.~P.~Blaizot and Y.~Mehtar-Tani,
  Nucl.\ Phys.\ A {\bf 929}, 202 (2014)
  doi:10.1016/j.nuclphysa.2014.05.018
  [arXiv:1403.2323 [hep-ph]].


\bibitem{Iancu:2014sha} 
  E.~Iancu and D.~N.~Triantafyllopoulos,
  Phys.\ Rev.\ D {\bf 90}, no. 7, 074002 (2014)
  doi:10.1103/PhysRevD.90.074002
  [arXiv:1405.3525 [hep-ph]].


\bibitem{Mueller:2016xoc} 
  A.~H.~Mueller, B.~Wu, B.~W.~Xiao and F.~Yuan,
  Phys.\ Rev.\ D {\bf 95}, no. 3, 034007 (2017)
  doi:10.1103/PhysRevD.95.034007
  [arXiv:1608.07339 [hep-ph]].


\bibitem{Baier:1996sk} 
  R.~Baier, Y.~L.~Dokshitzer, A.~H.~Mueller, S.~Peigne and D.~Schiff,
  Nucl.\ Phys.\ B {\bf 484}, 265 (1997)
  doi:10.1016/S0550-3213(96)00581-0
  [hep-ph/9608322].


\bibitem{CasalderreySolana:2007sw} 
  J.~Casalderrey-Solana and X.~N.~Wang,
  Phys.\ Rev.\ C {\bf 77}, 024902 (2008)
  doi:10.1103/PhysRevC.77.024902
  [arXiv:0705.1352 [hep-ph]].


\bibitem{Arnold:2009mr} 
  P.~B.~Arnold,
  Phys.\ Rev.\ D {\bf 80}, 025004 (2009)
  doi:10.1103/PhysRevD.80.025004
  [arXiv:0903.1081 [nucl-th]].


\bibitem{Idilbi:2008vm} 
  A.~Idilbi and A.~Majumder,
  Phys.\ Rev.\ D {\bf 80}, 054022 (2009)
  doi:10.1103/PhysRevD.80.054022
  [arXiv:0808.1087 [hep-ph]].


\bibitem{Luo:1993ui} 
  M.~Luo, J.~w.~Qiu and G.~F.~Sterman,
  Phys.\ Rev.\ D {\bf 49}, 4493 (1994).
  doi:10.1103/PhysRevD.49.4493


\bibitem{Collins:1981uw} 
  J.~C.~Collins and D.~E.~Soper,
  Nucl.\ Phys.\ B {\bf 194}, 445 (1982).
  doi:10.1016/0550-3213(82)90021-9


\bibitem{Kovchegov:1998bi} 
  Y.~V.~Kovchegov and A.~H.~Mueller,
  Nucl.\ Phys.\ B {\bf 529}, 451 (1998)
  doi:10.1016/S0550-3213(98)00384-8
  [hep-ph/9802440].


\bibitem{Majumder:2007hx} 
  A.~Majumder and B.~Muller,
  Phys.\ Rev.\ C {\bf 77}, 054903 (2008)
  doi:10.1103/PhysRevC.77.054903
  [arXiv:0705.1147 [nucl-th]].


\bibitem{Mueller:2012bn} 
  A.~H.~Mueller and S.~Munier,
  Nucl.\ Phys.\ A {\bf 893}, 43 (2012)
  doi:10.1016/j.nuclphysa.2012.08.005
  [arXiv:1206.1333 [hep-ph]].


\bibitem{Ghiglieri:2015ala} 
  J.~Ghiglieri, G.~D.~Moore and D.~Teaney,
  JHEP {\bf 1603}, 095 (2016)
  doi:10.1007/JHEP03(2016)095
  [arXiv:1509.07773 [hep-ph]].


\bibitem{Zakharov:2018rst} 
  B.~G.~Zakharov,
  JETP Lett.\  {\bf 108}, no. 8, 508 (2018)
  doi:10.1134/S0021364018200031
  [arXiv:1807.09742 [hep-ph]].


\bibitem{Dokshitzer:1991wu} 
  Y.~L.~Dokshitzer, V.~A.~Khoze, A.~H.~Mueller and S.~I.~Troian,
  Gif-sur-Yvette, France: Ed. Frontieres (1991) 274 p. (Basics of)


\bibitem{Wiedemann:2000za} 
  U.~A.~Wiedemann,
  Nucl.\ Phys.\ B {\bf 588}, 303 (2000)
  doi:10.1016/S0550-3213(00)00457-0
  [hep-ph/0005129].


\bibitem{Kang:2014ela} 
  Z.~B.~Kang, E.~Wang, X.~N.~Wang and H.~Xing,
  Phys.\ Rev.\ D {\bf 94}, no. 11, 114024 (2016)
  doi:10.1103/PhysRevD.94.114024
  [arXiv:1409.1315 [hep-ph]].
  
\end{thebibliography}
\end{document}